\documentclass[acmsmall]{acmart}
\AtBeginDocument{%
  }

\acmISBN{978-1-4503-XXXX-X/2018/06}

\setcopyright{cc}
\setcctype{by}
\acmJournal{PACMHCI}
\acmYear{2025} \acmVolume{9} \acmNumber{7} \acmArticle{436} \acmMonth{11} \acmPrice{}\acmDOI{10.1145/3757617}

\usepackage{subcaption}
\usepackage{caption}
\usepackage{fontawesome,dirtytalk}
\usepackage{xcolor}
\usepackage{graphicx,multirow,array}
\usepackage{color,soul}
\usepackage{tikz}
\usepackage{libertine}
\usepackage{rotating,tabularx}

\begin{document}

\title[It's a Complete Haystack: Understanding Dependency Management Needs in Computer-Aided Design]{It's a Complete Haystack: Understanding Dependency Management Needs in Computer-Aided Design}
\author{Kathy Cheng}
\affiliation{%
  \institution{University of Toronto}
  \city{Toronto}
  \country{Canada}}
\email{kathy.cheng@mail.utoronto.ca}
\orcid{0000-0002-6582-564X}

\author{Alison Olechowski}
\affiliation{%
  \institution{University of Toronto}
  \city{Toronto}
  \country{Canada}}
\orcid{0000-0001-5557-654X}

\author{Shurui Zhou}
\affiliation{%
  \institution{University of Toronto}
  \city{Toronto}
  \country{Canada}}
\orcid{0000-0002-6346-6073}

\renewcommand{\shortauthors}{Trovato et al.}

\begin{abstract}
In today's landscape, hardware development teams face increasing demands for better quality products, greater innovation, and shorter manufacturing lead times. Despite the need for more efficient and effective processes, hardware designers continue to struggle with a \textit{lack of awareness} of design changes and other collaborators' actions, a persistent issue in decades of CSCW research. One significant and unaddressed challenge is understanding and managing dependencies between 3D CAD (computer-aided design) models, especially when products can contain thousands of interconnected components. In this two-phase formative study, we explore designers' pain points of CAD dependency management through a thematic analysis of 100 online forum discussions and semi-structured interviews with 10 designers. We identify nine key challenges related to the traceability, navigation, and consistency of CAD dependencies, that harm the effective coordination of hardware development teams. To address these challenges, we propose design goals and necessary features to enhance hardware designers' awareness and management of dependencies, ultimately with the goal of improving collaborative workflows.
\end{abstract}

\ccsdesc[500]{Human-centered computing~Empirical studies in collaborative and social computing}
\ccsdesc[500]{Applied computing~Computer-aided design}
\keywords{awareness, collaboration, hardware development, distributed work}


\maketitle

\section{Introduction}
The technologies we rely on in our daily lives are constantly evolving, but the fundamental need for physical products remains unchanged. The process of designing physical products (i.e., hardware development) requires collaboration among multiple engineers and teams. In today's landscape, however, the hardware development process is increasingly challenging, due to the growing complexity of products and the distributed nature of hardware development work~\cite{cheng_age_2023,madsen_collaboration_2009}.

A significant challenge in collaborative hardware development is a \textit{lack of awareness}, a persistent issue highlighted in decades of CSCW research~\cite{mantau_awareness_2022,liu_let_2024} across various domains, like software development~\cite{treude_awareness_2010}, data science~\cite{mao_how_2019}, and collaborative writing~\cite{larsen_collaborative_2020}. Specifically in hardware development, it is challenging for designers to synchronize design activities and maintain an awareness of what others are working on, especially because products can be immensely complex. For instance, the Boeing 787 Dreamliner included 300,000+ CAD (computer-aided design) models~\cite{briggs_establish_2012} that must cohesively integrate for the final airplane to function properly. Even student design projects can involve 1,000+ models parametrically linked by 40,000+ dependency relationships~\cite{gopsill_emergent_2019}. Without dependency management tools, it is impossible to navigate these intricate dependency relationships~\cite{erhan_what_2020}, locate design errors~\cite{hahnlein_understanding_2024,ouertani_traceability_2011}, or prevent unintended impacts on others' work~\cite{cheng_age_2023}. Researchers, therefore, strongly advocate for the development of dependency awareness tools that can passively track these complex relationships~\cite{cataldo_identification_2006,klein_complex_2002,cheng_lot_2024}, thereby facilitating collaboration. 

Managing dependencies is not limited to hardware development; modern software development also requires careful coordination of activities, people, and artifacts. Software systems, like code, libraries, or frameworks, are built from existing systems, creating \textit{technical dependencies}~\cite{grinter_recomposition_2003,de_souza_supporting_2007}. Software developers must be aware of these dependencies to understand how their work will impact each other and to ensure that the project progresses in the right direction. Researchers have developed tools to analyze and visualize technical dependencies to decrease developers' cognitive load and facilitate communication among development teams~\cite{de_souza_supporting_2007}. 

While dependency management has been explored in software development, the unique nature of hardware development makes dependency management especially challenging, due to the geometric data, the granularity of CAD dependencies, and the lack of maturity in the tools. First, hardware designers work with 3D CAD models that represent intricate geometric and topological data~\cite{frazelle_new_2021}, and the models' dependencies are inherently linked to their physical architecture. Unlike software, where dependencies are typically expressed in text-based code and can be traced through direct imports or function calls, CAD dependencies are often implicit~\cite{cascaval_lineage_2023}, which are difficult to trace or search for~\cite{hahnlein_understanding_2024}. Second, dependencies in CAD can exist at multiple levels of granularity, such as between 2D sketches, or assemblies (composed of multiple interconnected 3D models), as well as between levels (e.g., a top-level assembly can reference a low-level sketch). Finally, collaborative infrastructure for CAD (e.g., awareness tools, version control) is generally less mature than in software development~\cite{cheng_age_2023}. Therefore, we can draw inspiration from the software development literature; however, we must recognize that the unique characteristics of hardware design necessitate tailored approaches to effectively manage CAD dependencies.



Dependency management tools tailored to CAD artifacts are crucial for supporting efficient collaboration in hardware development teams. While existing literature has developed visualizations or models of CAD dependencies~\cite{qamar_dependency_2012,masmoudi_dependency_2015}, the user requirements for such tools remain poorly understood due to the absence of a systematic exploration with real CAD practitioners. 
Therefore, we draw from similar research targeting awareness challenges in collaborative work~\cite{das_co11ab_2022,oleary_charrette_2018,head_managing_2019,wang_meeting_2024,kim_supporting_2021,abediniala_facilitating_2022}, conducting formative studies to better understand designers' needs regarding dependency management. Our work is guided by the research question: \textbf{What challenges do design engineers face with managing technical dependencies in CAD projects?} Gaining empirical insights into these challenges will enable us to devise appropriate strategies for addressing CAD dependency management, including the development of targeted tools.

Our approach to answering this question is a two-part formative study, wherein we gathered insights from actual CAD users through online forum discussions and semi-structured interviews. In the first phase, we mined 100 popular forum posts to obtain an overview of current practices and challenges. In the subsequent phase, we conducted interviews with 10 professional mechanical design engineers from a large technology organization that has established advanced dependency management practices, to identify rich scenarios and examples of persistent pain points. By conducting a thematic analysis of both data sources, we engage with a diverse array of CAD users through forum discussions, while simultaneously gaining an in-depth understanding of challenges from the interviews. This comprehensive approach enables us to thoroughly assess the existing gaps in CAD dependency management, as well as propose design implications for CAD platforms. Our contributions to CSCW are summarized as follows:

\begin{enumerate}
    \item A systematic identification of nine dependency management challenges in CAD, integrating insights from online forums and professional mechanical design engineers.
    \item A set of design goals, features, and initial tool concepts aimed at enhancing awareness and coordination in CAD work, focusing specifically on improving the traceability and navigation of technical dependencies.
    \item A discussion of the broader implications of improved technical dependency management, including its potential to enhance coordination of people and planning of design activities.
\end{enumerate}

\section{Background \& Related Work}\label{sec_background}

In this section, we first provide a background on awareness in CSCW to motivate our study. We then explore the specific awareness needs in CAD and the role of dependency management in addressing these needs. Finally, we review the relevant literature on CAD dependency management, as well as in the related field of software development, where dependency management has been studied more thoroughly. 

\subsection{Awareness in CSCW}\label{awareness}
%

Awareness is defined as the \textit{``understanding of the activities of others, which provides a context for your own activity''}~\cite{dourish_awareness_1992}, and is a fundamental concept to the field of CSCW~\cite{Gross2013,schmidt_problem_2002,leinonen_conceptualizing_2005,gutwin_group_2004}. In their seminal work, Gutwin and Greenberg~\cite{gutwin_descriptive_2002} developed a conceptual framework for \textit{workspace awareness}, which refers to the up-to-the-moment understanding of another person's interactions with a shared workspace. Simply put, awareness is understanding who is in the workspace, where they are working, and what actions they are taking~\cite{gutwin_effects_1999}, which involves both displaying one's actions and monitoring others' actions~\cite{heath_collaboration_1992}. More recently, Tenenberg et al. introduced \textit{we-awareness}, which extends beyond simply displaying one's actions and monitoring those of others~\cite{tenenberg_i-awareness_2016}. We-awareness refers to the socially recursive knowledge that each participant of a collaborative effort has of the other (i.e., the action of partner A is perceivable by partner B, partner B's perception is perceivable by partner A, and so on)~\cite{tenenberg_i-awareness_2016} such that all collaborators are mutually aware of each other's awareness~\cite{greenberg_implications_2016}. This recursive nature of we-awareness is crucial for collaborative work, especially in distributed teams, because it enables individuals to anticipate conflicts, coordinate activities, and build common ground~\cite{clark_grounding_1991}. 





There are various levels of collaborative work (\textit{coordinated}, \textit{cooperative}, \textit{co-constructive})~\cite{bardram_collaboration_1998}, but at the most basic level, collaboration requires coordination. As defined by Malone and Crowston, coordination is \textit{``the act of managing dependencies between activities performed to achieve a goal''}~\cite{malone_what_1990}. An example of a dependency is a \textit{transfer} dependency, which occurs when one activity produces something (e.g., a product, information) that another activity needs; this requires transferring from the ``producer'' activity to the ``consumer'' activity~\cite{malone_interdisciplinary_1994}. 
For coordination to occur, these dependencies must be managed, and the actors must have awareness of these dependencies. Without adequate awareness, teams are more likely to encounter misalignments, redundancies, or conflicts in their work~\cite{gutwin_group_2004}. Thus, awareness and coordination are deeply interconnected: awareness provides the contextual information necessary for managing dependencies, while coordination mechanisms~\cite{schmidt_cooperative_2011} aim to structure and handle those dependencies efficiently~\cite{duckert_collocated_2023,schmidt_organization_1994}.

Maintaining awareness, however, is challenging due to the dispersed nature of knowledge in collaborative work~\cite{gutwin_importance_2004} -- what Hutchins refers to as \textit{distributed cognition}~\cite{hutchins_cognition_1996}. The information needed for coordination is distributed across team members and their workspace. This dispersion imposes a cognitive load on individuals to track and process all relevant information~\cite{hollan_distributed_2000}. Researchers have developed various mechanisms (e.g., notifications~\cite{gutwin_usability_1996,lopez_awareness_2017}, shared dashboards~\cite{treude_awareness_2010}, real-time collaborative tools~\cite{yang_can_2023}) to reduce the user's cognitive load by providing visibility of other people's actions and changes to shared artifacts~\cite{dabbish_social_2012}. For instance, change awareness tools track asynchronous modifications to collaborative documents, summarizing what was changed, who made the change, where it occurred, how it differs from previous versions, and why it was made~\cite{tam_framework_2006}. Importantly, awareness mechanisms are not about delivering all information to users; rather, they are about disseminating contextually relevant information based on the specific work needs~\cite{schmidt_taking_1996}. Thus, designing effective awareness tools requires answering key questions like: \textit{What information should be provided? When should it be provided? To whom and in what form?}

Overall, the CSCW literature underscores the critical role of awareness and coordination in cooperative work. Our research specifically focuses on enhancing awareness in the field of hardware development by helping designers manage dependencies in CAD. By identifying CAD dependency management challenges, we highlight specific areas to improve designers' workspace awareness, ultimately fostering more effective coordination and collaboration for hardware development teams.

\subsection{Awareness Needs in Cooperative CAD Work}\label{sec: CAD awareness}

Awareness is key in all cooperative work domains, but it is especially difficult to maintain in CAD work, due to the nature of developing hardware products. For instance, CAD artifacts (e.g., 3D models) represent complex geometric and topological data~\cite{frazelle_new_2021}, making it challenging to use and develop traditional coordination tools, like version control~\cite{cheng_user_2023}. As such, CAD designers often struggle to identify the changes to a shared model, identify conflicts between models, and integrate contributions among collaborators~\cite{cheng_age_2023}.

In cooperative CAD work, multiple designers or teams work together to design a physical product. Typically, each designer or team is responsible for different parts or subsystems, which are later integrated to complete the overall product. To illustrate this process, imagine the following scenario: Designer A will model the wheels of a car, and Designer B will model the axle concurrently~\cite{asuzu_personas_2024}. For their designs to function cohesively, Designers A and B must communicate about design considerations, e.g., the wheel dimensions, materials, heat transfer, sequence of design decisions~\cite{pahl_engineering_2007}. Additionally, they must coordinate with Designer C, who may be designing the car's frame, and agree on the interfaces between components~\cite{sosa_misalignment_2004}. The CAD models act as intermediary objects that enable coordination between designers~\cite{tuikka_remote_2002, paavola_dynamics_2019}. Therefore, it is crucial for designers to be aware of their collaborators' changes to shared CAD models~\cite{cheng_age_2023}.

To ensure interfacing parts work together, designers use \textit{parametric CAD}~\cite{aranburu_how_2022}, which relies on parameters to define and constrain a model's geometry. Revisiting our car design example, Designer B may define the axle's position relative to the center of the wheels. If Designer A changes the wheel diameter, the axle will automatically update -- preserving fit and structural integrity without requiring manual rework~\cite{stark_major_2022}. While this approach works in theory, managing dependencies in complex products, comprising thousands of interacting parts, can quickly become a colossal task. Designers not only need awareness of the dependencies between CAD models to avoid unintentionally impacting another designer's model~\cite{cheng_age_2023}, but they also need to modify the dependencies accurately as the design changes -- for example, deleting dependencies when they are no longer needed. Maintaining accurate and consistent parametric relationships is essential to prevent costly manual rework and integration issues~\cite{camba_parametric_2016}. 

In summary, awareness in the CAD context entails not only understanding what others are working on but also recognizing how design changes affect the overall product. Due to the parametric aspect of CAD models, the dependencies within a product can become highly complex and, if not managed properly, can result in significant issues such as integration failures, misaligned components, and rework. Therefore, our work aims to shed light on the user needs for managing CAD dependencies, ultimately aiming to help improve awareness and coordination in CAD projects.



\subsection{Dependency Management}

In this section, we review related work, beginning with the study of dependency management in software development, a domain that has long recognized and supported these challenges, providing frameworks and tooling support. We then highlight how dependency management in CAD presents unique and significant challenges beyond those encountered in software development.

\subsubsection{In Software Development}

Dependency management has been studied widely in the software development field~\cite{de_souza_empirical_2008,de_souza_management_2003}.
In cooperative software development, dependencies arise among activities, artifacts, and different parts of the same artifact (like program dependencies~\cite{podgurski_implications_1989})~\cite{esbensen_routine_2014}. While there are many types of dependencies (e.g., knowledge, process, resource)~\cite{strode_dependency_2015,stray_dependency_2019}, software development projects generally have: \textit{technical dependencies}, which exist between artifacts (e.g., components of a software system~\cite{sun_code_2024})~\cite{de_souza_awareness_2011}, and \textit{work dependencies}~\cite{cataldo_socio-technical_2008}, among developers. Often, technical dependencies imply work dependencies~\cite{grinter_recomposition_2003} -- creating \textit{socio-technical dependencies}~\cite{sarma_tesseract_2009} -- which requires developers to coordinate and communicate to maintain a shared understanding of the dependency~\cite{grinter_recomposition_2003}.

Dependencies are necessary since modern software is built on the foundation of existing code, libraries, and frameworks~\cite{jahanshahi_beyond_2025}. However, dependencies can introduce challenges, such as breaking changes~\cite{jayasuriya_understanding_2023,brito_you_2019}, incompatibility with other dependencies~\cite{dietrich_dependency_2019}, and abandonment, where a package is used but no longer maintained~\cite{miller_understanding_2025}. Developers must keep updated with the latest versions of libraries that their code depends on to maintain \textit{dependency freshness}~\cite{cox_measuring_2015}, but it can be laborious to constantly track new updates~\cite{kula_developers_2017}, prompting the development of tools for automatic detection and updating~\cite{he_automating_2023}. When dependencies are not properly maintained, common errors can occur, such as missing dependencies (where necessary links are absent) and redundant dependencies (where a dependency exists but serves no purpose)~\cite{fan_escaping_2020,song_efficiently_2024}. Tools to automatically identify these errors have also been of interest in related work~\cite{fan_escaping_2020,omer_automatic_2011,rombaut_theres_2023}, and are widely used in open-source communities to facilitate the maintenance of dependencies~\cite{bogart_how_2016}.

Due to the complexity of managing dependencies, software development researchers have investigated various strategies, finding ``formal'' approaches, like impact analysis tools to identify entities that will be affected by a change~\cite{lehnert_taxonomy_2011}, and also ``informal'' approaches, like extensive email communication to broadcast the update~\cite{de_souza_management_2003}. Visualization tools have been built to help developers better understand the technical dependencies at different granularity levels, between files~\cite{gori_fileweaver_2020}, features~\cite{cho_feature_2008}, and code blocks~\cite{brown_facilitating_2023}. For instance, De Souza et al. created network graph visualizations to illustrate dependencies among developers for various scenarios, such as identifying developers working on similar tasks~\cite{de_souza_supporting_2007}. Another tool, Palantír, monitors artifacts in different developers' workspaces and visually exchanges information about how dependent artifacts are being modified, aiming to avoid conflicting changes~\cite{sarma_palantir_2012}. Gori et al. developed FileWeaver, a system that automatically detects dependencies among files, tracks their history, and lets users view the downstream and upstream dependencies~\cite {gori_fileweaver_2020}. The aim of all of these tools is to improve software developers' awareness of technical dependencies.


The software development literature has extensively explored various aspects of dependency management, such as types of dependencies, dependency-related errors, and awareness tools. However, managing dependencies in CAD is inherently more complex due to the nature of 3D models, which are tightly coupled with the physical architecture of the product, unlike text-based code~\cite{frazelle_new_2021}. Although software engineering provides mature frameworks and tooling for dependency management, comparable solutions remain underdeveloped in the CAD domain. Our work aims to fill this gap by providing a deeper understanding of dependency challenges in CAD, which is essential for developing effective management strategies. 

\subsubsection{In CAD}\label{bg:CAD dep management}


As introduced in Section~\ref{sec: CAD awareness}, the key idea of parametric CAD is that parameters are used to define all geometry, whether the shape of a single part, or the positions of parts relative to each other~\cite{gonzalez_facilitating_2024}. When designers are not careful about how they define these parameters (e.g., the direction of the dependency), they can introduce \textit{architecture debt}~\cite{rosser_technical_2021}, harming CAD model maintainability and reusability~\cite{camba_parametric_2016}. This becomes especially challenging in complex product development, where a single product (e.g., an airplane) may consist of hundreds of thousands of CAD models, all interconnected through intricate dependencies~\cite{briggs_establish_2012}. Consequently, there is a fundamental need for robust dependency management tools and strategies in CAD that can track relationships between different models, support impact analysis, and manage product variants~\cite{qamar_dependency_2012}.

Dependency management is particularly complex in hardware development due to three main factors: (1) the geometric nature of the data; (2) the granularity of dependencies; and (3) the lack of maturity in existing tools. In contrast to software -- where functions, variables, and modules convey explicit semantic meaning and clearly defined relationships -- CAD models are constructed through sequences of geometric operations or \textit{features}, that lack inherent semantics~\cite{cascaval_lineage_2023}. Features do not typically carry textual information about their design intent or functional role within the product~\cite{meltzer_whats_2024}. As a result, errors or changes are not as obvious; for example, a change could occur within the interior of a model, invisible from the outside view, whereas with software, code changes are more easily detected, highlighted, and summarized. This lack of visibility and semantic clarity makes it difficult to build tools that support effective change impact analysis in CAD. 

\begin{figure}[h]
  \centering
  \includegraphics[width=4.3in]{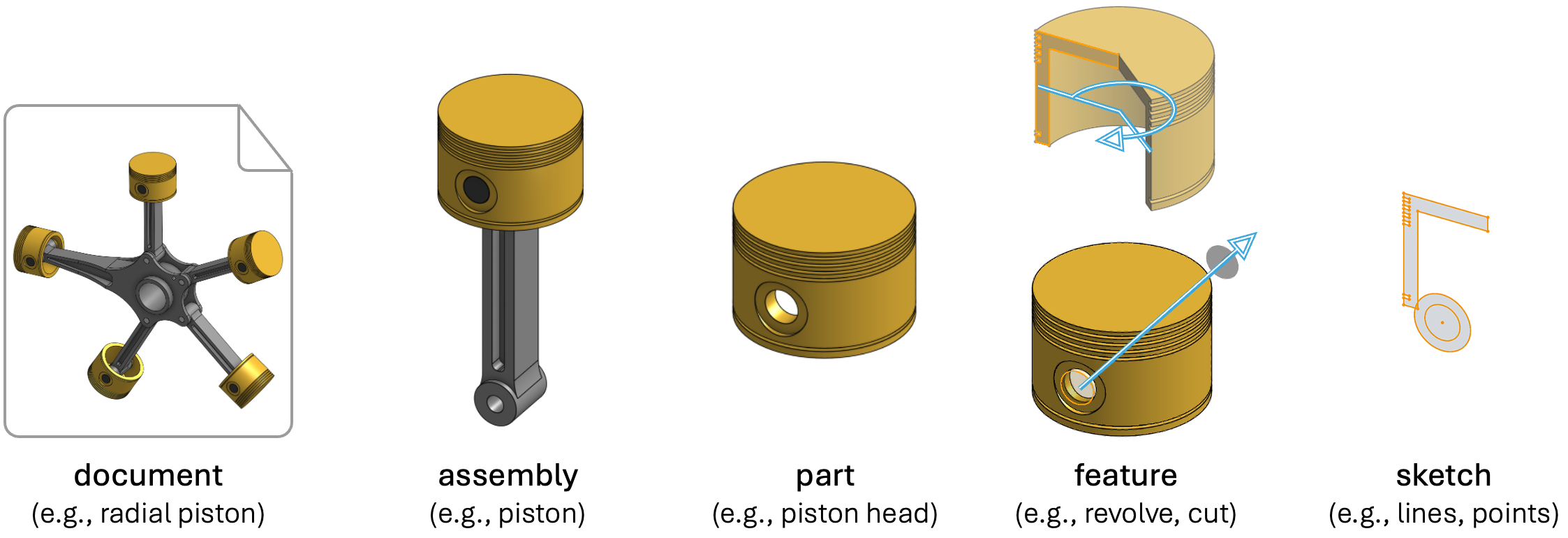}
  \caption{Granularity levels of CAD artifacts, organized from highest to lowest. A \textbf{document} is the top-level container for CAD artifacts and often represents a product. However, due to technical limitations in some CAD platforms, large or complex products may be split across multiple documents, creating ``external references.'' An \textbf{assembly} is a model composed of multiple interacting parts. A \textbf{part} represents a single physical object and is constructed from one or more features. A \textbf{feature} is an operation that manipulates geometry (in the figure, two features, \textit{revolve} and \textit{cut}, create the piston head part). At the lowest level, a \textbf{sketch} is a set of 2D line segments that defines a model's geometry. Models adapted from~\cite{demirkazik_onshape_2021}.}
  \Description{Granularity levels of CAD artifacts (e.g., document, assembly, part, feature, sketch).}
  \label{cadterms}
\end{figure}

Dependency management in CAD is further challenging because dependencies may span multiple granularity levels, ranging from low-level sketches and features to parts and assemblies (see Figure~\ref{cadterms}). These hierarchical relationships complicate both local edits and global coordination.

Furthermore, dependencies can exist between different documents, known as ``external references,'' a functionality supported by major CAD platforms, such as SolidWorks,\footnote{\url{https://help.solidworks.com/2021/english/SolidWorks/sldworks/c_External_References.htm}} Fusion360,\footnote{\url{https://www.autodesk.com/products/fusion-360/blog/reference-objects-sync-all-contexts/}} and Onshape.\footnote{\url{https://cad.onshape.com/help/Content/updating-references.htm}} External references can exist between artifacts at different granularity levels; for example, a circle sketch in one document could define the diameter of a pipe in another document, and any change to the sketch (e.g., increasing the diameter) will automatically update the dependent document.

\begin{figure}[h]
  \centering
  \includegraphics[width=3.2in]{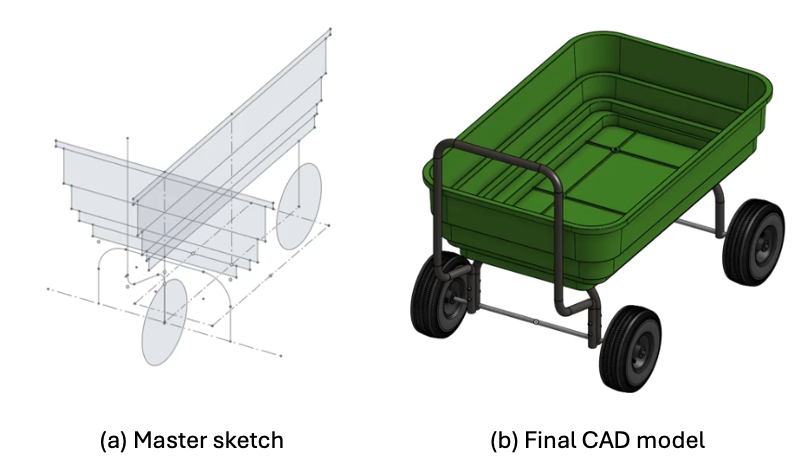}
  \caption{Complex products typically follow a \textit{top-down} architecture, where parameters are defined in (a) a master sketch, which is used to create (b) the final CAD model. Adapted from~\cite{onshape_tech_2022}.}
  \Description{Overview of findings, design goals, and features for a dependency management tool.}
  \label{mastersketch}
\end{figure}

It can be difficult for designers to trace the connections between different components~\cite{erhan_what_2020}, so best practices have been developed for systematically setting up dependencies. One common approach is ``top-down'' modelling~\cite{chu_multi-skeleton_2016,ciaccioli_investigation_2021}, whereby key parameters are defined in a central ``master sketch'', ensuring that dependent parts conform to the overall design constraints~\cite{pan_computer-aided_2016,chen_multi-level_2012}. Figure~\ref{mastersketch} illustrates this approach, where 2D sketches in the X-, Y-, and Z- planes define the reference geometries for the overall product. With master sketches, there is one central place where all of the dependency relationships are managed, and in the ideal scenario, changing the relevant element in the master sketch propagates the changes downstream accordingly~\cite{ciaccioli_investigation_2021}. 

Since multiple collaborators work across different dependent documents, researchers emphasize the importance of tracking dependencies between these external references. Existing efforts have developed automated measures for identifying these dependencies, such as through mapping links~\cite{masmoudi_dependency_2015}, applying network approaches to identify independent modules~\cite{sosa_misalignment_2004}, and using algorithms to detect redundant dependencies~\cite{farre_algorithms_2016}. Various researchers have also developed graph-based visualizations to help designers understand the dependencies between CAD models ~\cite{marchenko_new_2011,kozlova_graph_2011,johansson_supporting_2018}. Kozlova et al. interviewed six designers for feedback on their graph visualization of dependencies, finding that users desired interactivity, reduced visual clutter, and support for various levels of granularity~\cite{kozlova_graph_2011}. Despite these advances,
existing work focuses primarily on visualization and lacks support for managing the dependencies, such as enabling designers to modify dependencies directly within the interface. 
Beyond dependency management tools, awareness tools in CAD are generally less mature compared to software development. For instance, cloud-based architecture, though emerging, remains underutilized in many professional design organizations~\cite{Maher2017}, which limits collaboration. Sophisticated version control in CAD is still nascent~\cite{cheng_user_2023, cheng_lot_2024}, which hinders many aspects of dependency management, such as propagating changes across related documents or models~\cite{cheng_age_2023}. Given the current limitations of CAD collaboration systems, it is timely to guide the development of more advanced dependency management capabilities.


Existing research on CAD dependency management has introduced various methods to identify and visualize dependencies. However, a critical gap exists in understanding the challenges that must be addressed before effective solutions can be developed. With the exception of Kozlova et al.~\cite{kozlova_graph_2011}, most studies do not draw insights from real CAD users, and no studies take a systematic approach to understanding the problem. Our research aims to fill this gap by identifying user requirements for dependency management and proposing strategies grounded in empirical findings.

\section{Methods}

\begin{figure}[h]
  \centering
  \includegraphics[width=4.5in]{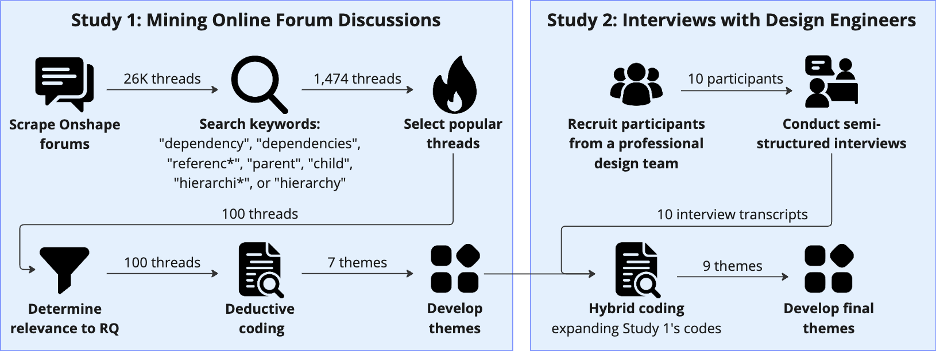}
  \caption{Overview of research methods for Study 1 and Study 2.}
  \Description{Overview of research methods for Study 1 and Study 2}
  \label{methods}
\end{figure}

To gain a comprehensive understanding of the challenges and user requirements surrounding CAD dependency management, we conducted an exploratory sequential two-phase formative study~\cite{Saldana2009}, and used Braun and Clark's reflexive thematic analysis (RTA) approach~\cite{braun_using_2006}. In Study 1, we mined and analyzed 100 user discussions from online CAD forum sites to build an initial understanding of the current challenges experienced by design engineers. This forum thread analysis revealed preliminary themes, which informed the analysis for Study 2, where we interviewed 10 professional mechanical design engineers. By following this multi-modal approach, we aim to overcome the limitations of a single data source. The forum discussions provide access to diverse discussion topics~\cite{cheng_user_2023} and a large pool of CAD users, enhancing the generalizability of our findings. In contrast, with interviews, we naturally reach fewer CAD users, but we gain in-depth insights that highlight the challenges and opportunities for CAD dependency management. Figure~\ref{methods} summarizes our research methods.

To determine appropriate sample sizes for both studies, we followed Malterud et al.'s guidelines for \textit{information power} in qualitative research~\cite{malterud_sample_2016}, which suggests that the number of participants required depends on the richness and relevance of the data rather than a fixed threshold. 
While either study alone may not hold sufficient information power, we conjecture that by supplementing and triangulating these two samples~\cite{wilson_triangulation_2006}, we collect the necessary evidence to answer our research question. 
We discuss further details on study design and sample sizes in the following subsections.

In both Study 1 and Study 2, we investigated dependency management in the context of \textit{Onshape},\footnote{\url{https://www.onshape.com/en/}} a cloud-based CAD and data management platform. Like all parametric CAD software, Onshape enables users to develop 2D sketches into 3D geometries and assemble these parts to create complex models. However, Onshape is particularly advanced in its support for dependency management for several reasons: (1) multiple designers can work on CAD documents simultaneously, removing restrictions on collaborative design; (2) the version control system enables branching and merging, supporting the creation and management of dependencies across different document versions; and (3) the cloud-based architecture eliminates common broken file path issues~\cite{stark_major_2022}. Given these capabilities, Onshape provides more sophisticated support than most CAD platforms, and is considered a state-of-the-art system for dependency management. Nonetheless, designers still report persistent challenges, therefore motivating the present study. Although we are interested in understanding dependency management in CAD overall, for this paper, we scoped our data collection to the Onshape platform, following the approach taken in similar studies, such as Kiani et al.'s investigation of help-seeking behaviours in the context of \textit{Autodesk Fusion360}~\cite{kiani_beyond_2019}. 


\subsection{Study 1: Mining Online Forum Discussions}\label{study1_methods}

The objectives of Study 1 are twofold: (1) to identify pain points and design goals for CAD dependency management tools, and (2) to guide the theme development for Study 2. To achieve this, we analyzed user discussions on online forums, a common data source used in CSCW research for understanding current practices and challenges from a wide range of perspectives~\cite{gilmer_summit_2023,almahmoud_vizdat_2023}. Online forums are important data sources for many domains, but especially for CAD-related topics, since they are one of the most frequently used resources among professional engineers and serious CAD users~\cite{kiani_beyond_2019,robertson_impact_2009,matejka_ip-qat_2011}. In our previous investigation of CAD version control practices~\cite{cheng_user_2023}, we found that forums can be considered communities of practice, offering rich discussions, user complaints, and needs that can inform the design of new features and improvements.

\subsubsection{Data Collection}
For this study, we collected data from Onshape forums,\footnote{\url{https://forum.onshape.com/}} which are officially affiliated with Onshape and hosted by its parent company, PTC Inc. At the time of our data collection (July 2024), Onshape forums comprised over 26K posts, over 121K comments, and over 10K members. Discussions on the forums fall into various predefined categories, with the largest being \textit{Community Support}, where users seek troubleshooting assistance, and \textit{Improvement Requests}, where users provide product feedback. 

To find forum posts relevant to our research question, we first did keyword searching using the following keywords (stemmed): \textit{``dependency''}, \textit{``dependencies''}, \textit{``referenc*''}, \textit{``parent''}, \textit{``child''}, \textit{``hierarchi*''}, OR \textit{``hierarchy''}.
We purposely did not stem the word \textit{dependency} to \textit{depend*} because it created a significant noise in the data, retrieving each instance of phrases like ``it depends''. Instead, to be comprehensive, we searched for both \textit{dependency} and \textit{dependencies}. We included the keyword \textit{referenc*} to capture discussions of external referencing, and \textit{parent}, \textit{child}, \textit{hierarchi*}, and \textit{hierarchy} as these are common terms used in the CAD context to describe artifact dependencies~\cite{camba_parametric_2016,cheng_analysis_2024}. 

To collect the forum post content, we used a custom web-scraping tool using Python, built with packages \textit{Selenium}\footnote{\url{https://www.selenium.dev/}} and \textit{BeautifulSoup}.\footnote{\url{https://pypi.org/project/beautifulsoup4/}} Using our keyword searching criteria, we retrieved any forum posts that contained one or more of these keywords in the title, initial post, or any comments. The data we collected includes each post's title, posting date, the content of the post, and the content in comments. Here, we define \textit{post} as the initial contribution by the original author and \textit{thread} as the full discussion, comprising the post and all subsequent comments.

In total, our keyword search yielded 1,474 threads. Given the heavy workload of manually coding all threads, we narrowed down the dataset by selecting the top 100 most popular threads, defining popularity by the number of comments, as these threads had the highest user engagement, and provided a rich source of content for analysis. The average comment count for popular threads was 63 (ranging from 35 to 401). The first author then reviewed each thread for relevance to our research question; irrelevant threads were tagged as such and excluded. When a thread was excluded, the first author revisited the original larger dataset ($n = 1,474$) and retrieved the next most popular thread, repeating this process until 100 relevant threads were identified. In the end, 14 threads were deemed \textit{irrelevant}, thus the total number of forum threads we reviewed was 114.

The decision to analyze 100 threads was guided by Malterud et al.'s concept of \textit{information power}~\cite{malterud_sample_2016}, which states that the larger the information power the sample holds, the lower the number of participants needed, and vice versa. A sample's information power depends on five dimensions: study aim, sample specificity,  use of established theory, quality of dialogue, and analysis strategy (case vs. cross-case). For our study, we ask an exploratory research question with a broad aim to understand CAD dependency management challenges, however, our study context is of a specialized field (of hardware development using mechanical CAD) with participants of a specific skillset. While forum posts lack direct interaction between researcher and participant -- reducing their information power -- they involve rich peer-to-peer dialogue, particularly in lengthy threads with multiple responses, where participants build on each other's contributions. To maximize information power, we prioritized discussions with the most comments, which are more likely to reveal diverse perspectives and rich ongoing exchanges. Across the 100 threads analyzed, there were 78 unique post authors and 841 unique members represented in the comments. We do not claim to capture complete insights on dependency management from all 841 users, but their excerpts indeed contributed to our theme development. Since the forums offer a diverse range of users, we need a bigger sample size (than with interviews) to provide sufficient information power. Given these considerations, we believe this dataset encompasses diverse perspectives from CAD practitioners, and that 100 forum threads provide sufficient data to \textit{``tell a rich story''}~\cite[p.~56]{clarke_successful_2013}.

\subsubsection{Data Analysis}
Our analysis followed Braun and Clark's method for reflexive thematic analysis (RTA)~\cite{braun_thematic_2021}. The strength of RTA lies in the active, \textit{reflexive} role of the researcher in the knowledge production process, where the researcher's subjectivity shapes theme development. In this work, the researchers are experts in modern CAD and software engineering, which is key to RTA, since it is our background that provides a unique perspective to investigate dependency management within the CAD context. During the thematic analysis process, the researcher's subjectivity deepens the interpretation of the data~\cite{braun_thematic_2021}. Our analysis approach was \textit{inductive}, given the exploratory nature of our research question. We focused on both \textit{semantic} and \textit{latent} levels to reflect the explicit content of the data while leveraging our expertise to inform our interpretation of the data.

There are six phases of RTA, which follow a recursive process~\cite{clarke_successful_2013}. The first phase, \textit{familiarization}, began with the first author reading through all 100 threads to determine their relevance to the research question, making casual observational notes throughout. The next phase, \textit{generating codes}, involved assigning short descriptive labels or ``codes'' to data excerpts to identify themes without following a predefined framework, resulting in 290 codes. In the third phase, \textit{constructing themes}, we collated similar codes to develop themes; e.g., a recurring theme was that CAD designers struggled to anticipate how changes to an artifact would affect its dependents. Throughout the theme construction phase, all authors iteratively refined the candidate themes to ensure that each was distinct and had a central organizing concept, resulting in 14 candidate themes and 73 candidate sub-themes. In the fourth and fifth phases, \textit{revising} and \textit{defining themes}, all authors collaboratively reviewed each theme's codes, coded data, and relation to the research question to identify the most meaningful themes. We then defined the following candidate themes: \textit{difficulty in tracing dependency chains}, \textit{lack of overview of project structure}, \textit{poor impact analysis}, \textit{ambiguous dependency freshness}, \textit{difficulty reorganizing models within the hierarchy}, \textit{broken dependencies}, and \textit{disorganized design history}. These seven themes provided an initial framework for the thematic analysis of the interview transcripts generated in Study 2, detailed in the following section.

\subsection{Study 2: Interviews with Design Engineers}\label{study2_methods}
For our second formative study, we conducted 10 semi-structured interviews with professional mechanical design engineers who use CAD. The semi-structured format allowed us to ask open-ended questions that captured a broad range of topics, while also providing an opportunity to triangulate the themes identified in Study 1. 
The aim of Study 2 is three-fold: (1) to validate the findings from the forum analysis; (2) to gather rich scenarios and nuances that highlight why dependency management remains a persistent issue in CAD, and (3) to gather feedback on tool features that could improve engineers’ awareness and management of dependencies. 



\subsubsection{Participants}
We recruited participants from a large technology organization that produces software and hardware products -- for anonymity, we refer to them hereafter as \textit{Company~X}. While Company~X has a substantial workforce with many departments and engineering teams, we focused our participant recruitment on a particular team (hereafter: \textit{Team~Y}) that specializes in the design and development of autonomous robots. We chose this particular company and team for a few reasons. First, Company~X is one of the largest organizations (>10K employees) that use Onshape, with a substantial workforce and a diverse range of complex products. This complexity requires sophisticated workflows and robust data management strategies for effective coordination, which allows us to draw more meaningful insights for studying dependency management. Second, Team~Y has established a significant history of using Onshape for over four years. Consequently, their practices regarding regular Onshape usage are well-established. Finally, Team~Y, makes extensive use of the external referencing features, relying on these workflows to design their flagship robot; thus, we are capturing persistent challenges with using state-of-the-art dependency management tools, not challenges due to a lack of familiarity. All of these factors make Team~Y an ideal pool to recruit participants to understand dependency management challenges. 

We used purposive sampling~\cite{Campbell2020} to recruit participants from Team~Y. We recruited online through a screening questionnaire posted in Team~Y's relevant Slack channels. In total, we interviewed 10 participants (represented by ID1-ID10 in Table~\ref{participant_table}), of which four were women, and six were men. All interviewees have at least two years of experience working in Team~Y at Company~X and using Onshape professionally. All participants were experienced in multiple CAD softwares, and nine had experience using CAD software other than Onshape in a professional setting. Five participants were senior mechanical engineers, three were mechanical design engineers, and two were team leads. Each member of Team~Y works on all the subsystems within the robot, so each interview participant has experience managing dependencies from both sides of the dependency relationship, which was an important experience for our interviews. 

In determining a suitable sample size, we used Malterud et al.'s \textit{information power}~\cite{malterud_sample_2016}. We purposefully selected interview participants based on specific characteristics of the organization, team, and individual, to ensure that their insights were highly relevant to our research question. As researchers with extensive experience conducting interviews and studying CAD collaboration, we engaged in in-depth discussions with participants, further enhancing the quality of dialogue. Additionally, because all interview participants belonged to the same company, a smaller sample size was sufficient for identifying patterns within this organizational context. Throughout recruitment and data collection, we continuously assessed the relevance and quality of the data to appraise the information power iteratively, as recommended by Malterud et al.~\cite{malterud_sample_2016} and Braun and Clark~\cite{braun_thematic_2021}. Based on this ongoing evaluation, we determined that 10 participants provided sufficient information power, and concluded data collection at that point.




\begin{table}[h]
\footnotesize
  \caption{Participants in Study 2. Columns with Yrs. indicate the number of years of experience.}
  \begin{center}
  \label{participant_table}
  \begin{tabular}{l c c c c c}
     \toprule
    Participant&Job Title&Gender&Yrs. using CAD&Yrs. using Onshape&Yrs. at Company\\
    \midrule
    ID1&Senior Mechanical Engineer&M&7 - 9&1 - 3&1 - 3\\
    ID2&Senior Mechanical Engineer&M&10 +&1 - 3&1 - 3\\
    ID3&Senior Mechanical Engineer&M&10 +&4 - 6&4 - 6\\
    ID4&Mechanical Design Engineer&M&10 +&1 - 3&1 - 3\\
    ID5&Senior Mechanical Engineer&M&10 +&4 - 6&4 - 6\\
    ID6&Mechanical Design Lead&M&10 +&1 - 3&7 - 9\\
    ID7&Engineering Team Lead&W&10 +&4 - 6&4 - 6\\
    ID8&Mechanical Design Engineer&W&1 - 3&1 - 3&1 - 3\\
    ID9&Mechanical Design Engineer&W&4 - 6&1 - 3&1 - 3\\
    ID10&Senior Mechanical Engineer&W&7 - 9&4 - 6&4 - 6\\
   \bottomrule
  \end{tabular}
\end{center}
\end{table}

\subsubsection{Procedure}

All 10 interviews were conducted remotely via Zoom between August and September 2024. The sessions were audio- and screen-recorded and automatically transcribed using Zoom's transcription service. Each interview lasted an average of 57 minutes, ranging from 47 to 65 minutes. The first author conducted all the interviews. 

We began by asking participants about their role, job responsibilities, and experience at their organization. We refrained from asking about specific challenges to avoid biasing their responses based on themes identified in the forums. Instead, we asked them to describe a recent design project and walk us through the entire process -- from starting the CAD work to delivering the final output. We encouraged participants to provide examples to ground their responses. For each scenario they described involving the management, modification, or understanding of dependencies in CAD, we asked follow-up questions, such as what artifacts were dependent, what information they needed to understand the dependency relationships, and what challenges they faced. As participants recounted scenarios, they shared their screens while using Onshape, walking us through the features they typically use and demonstrating their design and navigation processes. 
The study was approved by the University of Toronto's Ethics Review Office.

\subsubsection{Data Analysis}
To begin data analysis, the interview transcripts were downloaded from Zoom, anonymized, and imported into the qualitative data analysis platform, NVivo-12.\footnote{\url{https://lumivero.com/products/nvivo/}} 
Following Study 1's approach, we conducted the six phases of reflexive thematic analysis (RTA)~\cite{braun_using_2006} and coded both semantic and latent levels of the data. However, we employed hybrid coding to analyze the interview transcripts, combining deductive and inductive coding~\cite{Corbin2008}. Like in Study 1, the first author began with the \textit{familiarization} phase by reviewing and cleaning the automatically transcribed data while noting initial ideas. Since the data from Study 1 had already been coded, the first author marked instances where interview data aligned with existing candidate themes, along with new ideas that stood out against the previous analysis. 

During the \textit{generating codes} phase, the first author systematically analyzed each interview, generating around 360 codes. In the \textit{constructing themes} phase, we built upon the seven candidate themes of dependency management challenges developed in Study 1, organizing codes that matched these existing themes accordingly. When we identified new challenges related to dependency management, we created additional themes to capture these insights. The first author independently coded the interview transcripts, while all authors collaboratively discussed and refined the themes throughout the process. This iterative refinement involved comparing the themes against each other and checking them against the original codes and dataset, per the Braun and Clark method~\cite{braun_using_2006}. At this stage, the thematic analysis included 20 candidate themes and 57 candidate sub-themes. 

During \textit{revising} and \textit{defining themes}, all authors examined the codes within each theme to confirm they follow a coherent pattern, and revisited the dataset to determine that the themes accurately reflect the participants' original meanings. This interview analysis confirmed five of the seven themes from the forum threads and found two additional themes: \textit{messy navigation of the master sketch} and \textit{dependency conflicts across versions}. Across both formative studies, we identified nine challenges of CAD dependency management. 

Finally, in the last phase of RTA, \textit{writing the report}, we developed a deeper sense of how the themes fit together to create a cohesive picture of CAD dependency management challenges. Through this process, all authors collaboratively developed overarching themes that group the nine themes into challenges related to: (1) traceability; (2) navigation; and (3) consistency. 

\section{Findings: Challenges of CAD Dependency Management}\label{results}

This section synthesizes the findings from Study 1 and Study 2. We found nine challenges that design engineers face with managing technical dependencies in CAD projects, summarized in Table~\ref{results_table}. To present our results, we have grouped the challenges into three overarching themes: traceability-, navigation-, and consistency-related. Quotes from forum threads are denoted with \textit{F} (for ``forum'') and the year it was posted in brackets -- for example, ``(F24)'' for a thread posted in 2024. Quotes from interviews are denoted with the participant ID -- for example, ``(ID1)''. 

\begin{table}[h]
\footnotesize
  \caption{Overview of CAD dependency management challenges, as identified through thematic analysis of forum threads and interviews conducted in Study 1 and Study 2, respectively. We summarize the frequency of each challenge based on the number of relevant forum threads (out of 100) and the number of interviews (out of 10) that reported these challenges. A dash (---) indicates that the challenge was not identified.}
  \begin{center}
  \label{results_table}
  \begin{tabular}{l l c c}
     \toprule
    Category&Challenge& \# of Threads & \# of Interviews\\
    \midrule
    Traceability-related&Difficulty in tracing dependency chains & 36 & 10 \\
    &Poor impact analysis& 19 & 8 \\
    &Broken dependencies& 11 & --- \\
    Navigation-related&Lack of overview of project structure & 23 & 4 \\
    &Difficulty reorganizing models within the hierarchy & 5 & 2 \\
    &Disorganized design history& 8 & --- \\
    &Messy navigation of the master sketch & --- & 5 \\
    Consistency-related&Ambiguous dependency freshness & 13 & 4 \\
    &Dependency conflicts across versions& --- & 1 \\
   \bottomrule
  \end{tabular}
\end{center}
\end{table}

\subsection{Traceability-related}
Traceability-related challenges focus on the difficulty of understanding how different artifacts are interconnected, including tracing the origins and histories of designs. Additionally, inadequate traceability hampers designers' ability to understand how artifacts are impacted by changes.

\subsubsection{Difficulty in tracing dependency chains}\label{chal:tracing}

Before beginning any modelling work -- whether modifying an existing design or creating a new one -- CAD designers need to know how design artifacts are interconnected. This requirement is central to parametric CAD~\cite{ganeshan_framework_1994}, which requires thoughtful consideration of how dimensions drive the overall design. However, tracking all these relationships becomes challenging in complex product development, where products consist of numerous parts and subsystems with intricate dependencies. As one forum user expressed, \textit{``external references are problematic to humans, because we do not, and can not, scan through everything we ever knew which bears on the problem at hand, every time we make decisions''} (F15).

\begin{figure}[h]
  \centering
  \includegraphics[width=4in]{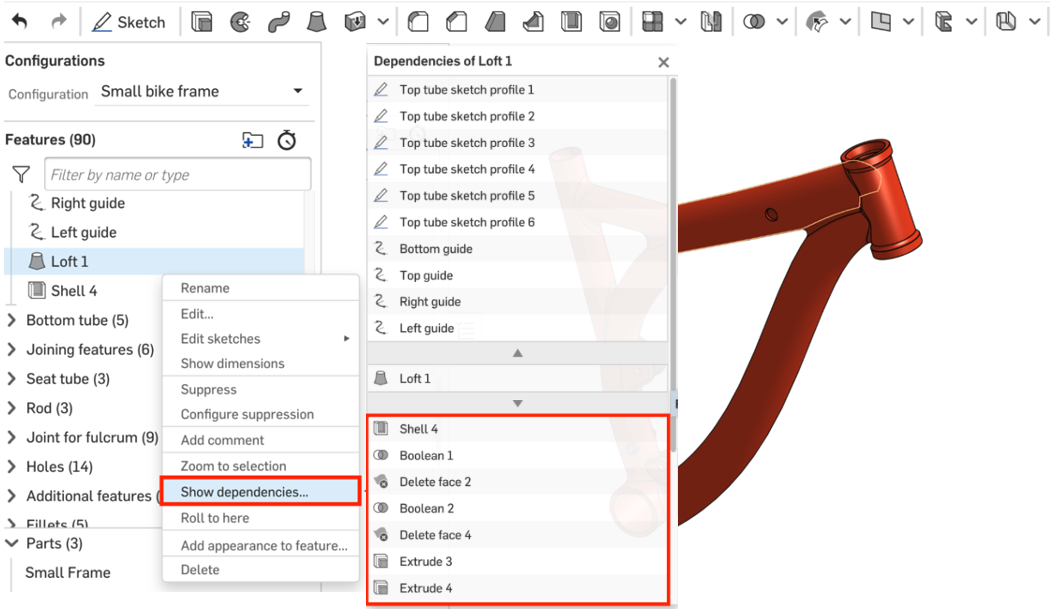}
  \caption{Onshape's ``Show dependencies'' feature enables traceability within a document. Adapted from~\cite{onshape_show_2022}.}
  \Description{Overview of findings, design goals, and features for a dependency management tool.}
  \label{showdependencies}
\end{figure}

Existing tools typically only allow users to trace dependencies for one feature at a time (see Figure~\ref{showdependencies}). Users must manually select a feature, inspect its dependencies, and then repeat this process for each related feature. Currently, there is no way to generate a comprehensive list that spans multiple levels of granularity (e.g., across features, parts, assemblies).
This process becomes especially challenging when several layers of hierarchy exist (as described in Section~\ref{bg:CAD dep management}), since dependency tracing is only possible within a single document. As a result, designers must trace the model's history across multiple documents to understand the design intent. Often, this involves discovering that one model depends on another in a separate document, which itself may depend on yet another model, leading to a laborious and manual tracing process. As interviewee ID5 explained, \textit{``you could track something from one part to the next part, and you will go into Master Sketch, and that Master Sketch derived another Master sketch, etc. [...] You would have traversed 8 different places before you actually found out where the change was.''} Participants identified two primary scenarios in which they perform dependency tracing: (1) to understand a model's history before making modifications, and (2) to investigate the root cause of a design error. It is important to note that both scenarios involve tracing \textit{upstream} across documents, whereas it is \textit{``almost impossible to track downstream''} (ID2). ID2 further explained that if a tool could list all downstream features linked to a sketch, this added visibility would influence their design decisions. For example, \textit{``if [the sketch] points to a million different stuff, then probably I'll leave it alone, or be really, really careful''} (ID2). 

The difficulty of tracing dependency chains was the most common challenge, mentioned in 36\% of forum threads and by all 10 interviewees. 

\subsubsection{Poor impact analysis}\label{chal:change propagation}
Impact analysis is understanding how changes to a parent artifact propagate downstream, which, in CAD, is often a difficult and manual task. Onshape -- compared to other CAD platforms -- improves this process by allowing change propagation for models within the same document, as one interviewee noted, \textit{``if the part geometries were just defined in the same document, you will never have to do [change propagation]''} (ID5). However, a significant challenge remains: while the software can manage change propagation for a single dependency layer, it cannot efficiently handle multi-level hierarchies that span across multiple documents. As interviewee ID1 said, \textit{``it takes a lot of time and effort to cascade a change down into the hierarchy. It might be 5 levels deep or something.''} Although manually reviewing and approving each propagated change in every affected document is laborious, some users find this stepwise process purposeful. One forum user explained that it allows designers to manually check each modified document \textit{``to ensure that each parent assembly is referencing the correct/latest released version of a child part''} (F23).

Given the tedious nature of this task, it can be impractical for designers to verify the impact in each document, which can cause unintended changes to models without the designer's awareness. Interviewee ID3 described a common scenario: \textit{``every time there is an update of the master sketches, you need to go to all the documents, update that reference, and then cross-validate.''} Without tools to systematically track these impacts, engineers rely on heuristics and a deep understanding of their design to navigate dependencies selectively, since checking every child artifact for changes is impractical. As ID3 further emphasized, \textit{``you need to be a little bit selective. But that is based basically [based] on the knowledge that we have. We don't have a bulletproof method here. It's just too much.''}

The idea of automatic change propagation is appealing to streamline this process, but users have expressed caution about this approach. One forum user stated, \textit{``I'm interested in the possibility of automated version updating in the assemblies, though I suspect that will get me in big trouble when at some point when I forget to check an assembly that's been auto-updated''} (F21). With or without automation, users highlight the importance of having a real engineer review the changes, stating, \textit{``don't forget that the user is the ultimate judge of their intent. Please make sure we can override any magic if the magic isn't working''} (F15). Overall, CAD users find manual change propagation tedious, while fully automated propagation can be prone to errors. Thus, the ideal solution may be to streamline the review and approval process to improve efficiency, while maintaining design integrity by systematically identifying changes. 

The challenge of poor impact analysis was mentioned in 19\% of forum posts and by 8 interviewees.

\subsubsection{Broken dependencies}\label{chal:broken}

As part of refactoring practices, designers may clean up or archive old designs without realizing these are still linked to other parts, unintentionally creating broken dependencies. One forum user shared a typical scenario: \textit{``I have a document with a sketch [derived] from another document which I trashed thinking it was not needed. [...] It really seems like this shouldn't be possible, at least without a warning''} (F22). In this case, the designer was unaware of the dependency and received no warning before creating the error.

Another frustration users face is the lack of detailed information about the root cause of the error. For instance, Onshape's generic message \textit{``Some dependencies are missing''} leaves users uncertain about how to fix the issue. One user drew the comparison, \textit{``This is a bit like a compiler getting to the end of the compilation and saying, `Your source program has some errors. Please fix them.' Not useful''} (F22). Without knowing what the dependency originally referenced, designers cannot repair or reroute the relationship to another artifact.

The challenge of broken dependencies was mentioned by 11\% of forum threads but by no interviewees. 

\subsection{Navigation-related}
Navigation-related challenges involve the efficient retrieval and organization of information. These challenges often arise from disorganized data, requiring designers to invest significant time and effort to make sense of the cluttered dependency information.

\subsubsection{Lack of overview of project structure}\label{chal:overview}

During CAD modelling, designers often have multiple documents open to review the relevant information, but they typically focus on design work within a single document at a time. Working predominantly within one document can create silos, making it challenging for designers to understand how the design of the product is progressing at a high level; this is especially the case when multiple collaborators are making changes to different subsystems that may not immediately impact one another. Consequently, the inability to see the bigger picture of these interconnected artifacts presents a significant challenge. One forum user expressed, \textit{``with design information distributed between documents, branches, versions, revisions [...], it's very hard to look at a project and quickly get an understanding of what's going on''} (F19). 

\textbf{Feature request: visualizing external references.} Designers have suggested that a visualization would be useful to address the lack of overview problem. Interviewee ID9 stated, \textit{``the only way for us to find out how [designs are] used is really by right-clicking and ``Show dependency''. It's all text information [...], and we actually don't have a visualization about how documents are linked together''} (ID9). This sentiment was echoed in the forum posts, where one user wrote, \textit{``I think a visual representation of the files makes it easier for a visual program''} (F14), indicating that such a tool would align with the way design engineers think and work, given the inherently visual nature of CAD. Additionally, since the products being designed are physical, some users expressed that the dependency structure should be \textit{``ideally based on the assembly structure''} (F19), mirroring the architecture of the intended product. 

This lack of overview challenge was mentioned in 23\% of forum posts and by four interviewees.

\subsubsection{Difficulty reorganizing models within the hierarchy}\label{chal:hierarchy}
As external references are created between CAD documents, these documents form a hierarchical structure. During the initial design phases, designers often struggle to determine where certain dimensions or sketches should be placed within the hierarchy; for example, in our previous car design scenario (Section~\ref{sec: CAD awareness}), designers need to decide the hierarchy of the axle, wheel, and frame dimensions. ID5 explained, \textit{``if I have one Part [A] here and another Part [B] here, and then they have some sort of relation in common, then I don't want to be designing this [Part A] in isolation and this [Part B] in isolation. I want them to have a common ancestor, a common source of truth.''} To achieve this, the sketch that defines the interaction between the two documents is moved to an upper-layer document, so that any change will propagate to both lower-layer documents. This approach works in theory, but as the design progresses, it can become necessary to reorganize the hierarchy. For example, lower-layer dimensions might need to be moved higher if they will now drive multiple documents, or vice versa, as ID5 noted: \textit{``if a sketch is no longer needed, then we shouldn't define that in an upper document, because that just creates an unnecessary step.''} However, interviewees expressed that changing the hierarchy is not that simple, because, \textit{``when moving the sketches, everywhere these are referenced, you have to reintroduce that reference somewhere''} (ID4). For instance, in our car example, if the frame depends on the axle, which depends on the wheel, and now the frame needs to be the driving model, designers must carefully adjust all related references. Identifying which dependency relationships need to be edited, removed, or rerouted is not obvious, and there is no systematic or automated way to inform the user of which references require changes.

Similarly, interviewees described difficulties in decomposing one large document into smaller ones. ID9, the designer who originally developed the master sketch architecture, explained: \textit{``it was a pain to split this master sketch [...] into 3. We needed to think about how we define the boundaries.''} This process becomes more error-prone when done collaboratively, as multiple people may be rerouting these dependency relationships across various product subsystems. Without full awareness of each other's actions, errors such as circular references can occur. ID9 warned: \textit{``you need to plan through like which part is dependent on which and how to avoid a loop reference because it's really easy to chase your own tail when you all work together.''}

In both scenarios -- reorganizing the hierarchy or decomposing documents -- moving references must be handled carefully to avoid errors when the references are not properly redefined. This challenge was mentioned in 5\% of forum posts and by two interviewees.


\subsubsection{Disorganized design history}\label{chal:history}

In CAD, the feature tree is a hierarchical representation of all the features in a 3D model or assembly, reflecting the design history of the CAD document. As designs become more complex, designers report increasing difficulty navigating the tree. One forum user explained that when \textit{``starting to deal with bigger part studios (Still relatively small at 65 parts), [they are] really feeling the need for list organization''} (F14). A common frustration is the inability to easily reorder features so that they are displayed in a more intuitive order. For example, designers expressed the desire to group dependent features. As one user explained, \textit{``I often want to reorganize my tree so that features are as far up as they can be without breaking. I'd be so happy if I could right-click a feature and just pick `Move Below Next Parent,' and it would push it as far up as possible in the tree''} (F18). The ability to reorganize the feature tree would provide designers with a clearer overview of how certain features relate. One user highlighted the need for collapsing certain features together, stating, \textit{``one of the main reasons for organizing a feature tree via grouping (like folders) is to condense it so that the sequence of actions can be more easily understood by a newcomer or your future self''} (F15). 

This challenge was mentioned in 8\% of forum threads, but did not appear in the interviews.

\subsubsection{Messy navigation of the master sketch}\label{chal:mastersketch}

In complex product development, designers often employ a ``master sketch'' architecture to define the critical dimensions used throughout the product (as described in Section~\ref{bg:CAD dep management}). However, as more sketch elements accumulate, the master sketch can quickly become overwhelming and difficult to read. ID10 captured this issue, stating, \textit{``there's about a billion things. And there's dots and lines over everything. It's a complete haystack, [and] very rapidly, because we're working in 3 dimensions, [the master sketch] becomes unreadable.''} This messiness of the master sketch not only hinders effective navigation but also increases the risk of errors. For example, densely packed sketches make it challenging to select the correct reference geometry, as ID9 noted: \textit{``if two lines are 3mm apart from each other, it's easy to get things wrong.''} 

To improve readability, designers often want to \textit{``clean it up''} (ID10) by reorganizing or removing unnecessary sketch elements. One common scenario is that \textit{``some sketches, like lines [or] reference points, are no longer used, as the design changes and updates''} (ID9). However, Onshape does not alert the user that a sketch element is no longer needed and can be safely removed. ID9 described their two options: \textit{``the first way is to read it, think about it, and understand it, which takes so much time that it's impossible. The second way is just to delete it and see which document goes red [i.e., has an error], which is also not very optimal.''}

If designers are unaware of a sketch's downstream dependencies, modifying or removing it can lead to unintended consequences (Section~\ref{chal:change propagation}). As ID10 explained: \textit{``If I delete [a sketch element], I'll break [all these] parts, and I won't know about it until I update all references. And if I don't know where those references are, I can't fix them. Someone might find that out 2 months later when they update for some completely other reason, and then everything breaks, and they don't know why.''} This uncertainty makes designers hesitant to delete anything for fear of causing broken dependencies. As ID4 said, \textit{``stuff stays in [the master sketch] for a lot longer than it needs to, because the safe thing to do is not to delete it. You don't know what chaos you'll cause if you get it wrong.''} Consequently, outdated features remain in the sketch, further adding to the clutter and increasing the risk of accidentally selecting an outdated element to reference later on.

The challenge of messy master sketches was significant, mentioned by 5 out of 10 interviewees.

\subsection{Consistency-related}
Consistency-related challenges arise when versions of dependent documents are not synchronized. This leads to two main issues: designers may be unsure whether dependencies are up-to-date (i.e., the freshness of the dependency), and conflicting versions can occur.

\subsubsection{Ambiguous dependency freshness}\label{chal:ambiguity}
External references in CAD are intended to allow parts and subsystems to be developed in parallel, facilitating collaboration and design efficiency. As design work progresses, new versions of documents are created, and dependent artifacts need to be updated to ensure changes transfer correctly across documents. When a document has a new version, Onshape sends notifications to all downstream documents, indicating that the dependency is outdated and requires updating. However, it is often unclear whether the change actually impacts other designs. This ambiguity becomes particularly challenging in the CAD context because each document may contain various artifact types (e.g., sketches, features, parts, assemblies), and a change to any one artifact will trigger a new version for the entire document -- even though the specific dependent artifact did not change. Assessing the relevance of this `new version' notification across numerous dependent documents is a time-consuming and complex task. One forum user described this frustration: \textit{``I noticed that ANY change to information referenced in a drawing, whether the change affects what appears on the face of the drawing or not, requires the drawing to be updated. [...] I spend a lot of time hitting the `update' button and waiting for the drawing to regenerate''} (F19). 

To avoid missing critical updates, some designers err on the side of caution, creating new versions and updating references frequently. However, this approach has its drawbacks, as another forum user explained: \textit{``Let's say I have a Document with 10 tabs and I change one of them. I'm happy with the change, so I create a new version... well what just happened is that I just versioned 9 Elements that DID NOT CHANGE. To me, that is not a good Data Management practice and creates a lot of overhead keeping track of what's changing and what's not''} (F19). 

Beyond substantive design changes, even minor metadata updates in a document can trigger notifications, further adding to the noise. As interviewee ID1 observed, \textit{``Onshape is meant to only signal version updates when they have an impact. But, Onshape's definition of an impact might differ from what we think. [The software] might think that changing some description is a version change.''} This notification system's ambiguity and the lack of granularity in alerts make it challenging for designers to distinguish between meaningful updates and unnecessary distractions.

The challenge of ambiguous dependency freshness appeared in 13\% of forum posts and four interviews. Due to the frustration this issue caused, Team~Y developed an in-house tool to automate this version update process. This tool generates a list of all ``out-of-date'' dependencies for any document, enabling users to select and update these references in bulk, which significantly streamlines the process and \textit{``kills all the noise of the blue dots [i.e., update icons]''} (ID4).

\subsubsection{Dependency conflicts across versions}\label{chal:conflict}

Dependency conflicts in CAD arise when documents reference different versions of another document. If dependencies fall out of sync, design components can become incompatible. 
For example, ID1 highlighted, \textit{``if I [reference] revision B in a product and revision F in a later product, how do you actually ensure that they work together?''} Moreover, this issue extends beyond the CAD environment to physical, real-world products. ID1 noted that \textit{``we're gonna have multiple versions of the bot running around [...] with multiple references to these master sketches. A physical product out there won't ever have a trigger to update, right? So how do we keep track of all of those interweaving versions?''} The challenge becomes even more complex when trying to maintain consistency across various product lines, each tied to different states of the same master sketch.

This challenge was only mentioned by ID1 and did not appear in any forum discussions; however, we include it in our findings because of the significant negative consequences of version conflicts.

\section{Discussion}

Here, we summarize the identified challenges and outline key design goals for a CAD dependency management tool. We then present potential features, design mockups, and a user scenario to demonstrate how the proposed tool can address these challenges. Additionally, we discuss our study's main implications, concluding with limitations and future work directions.

\subsection{Design Implications for CAD Dependency Management Systems}\label{sec:design goals}



\begin{figure}[h]
  \centering
  \includegraphics[width=4.7in]{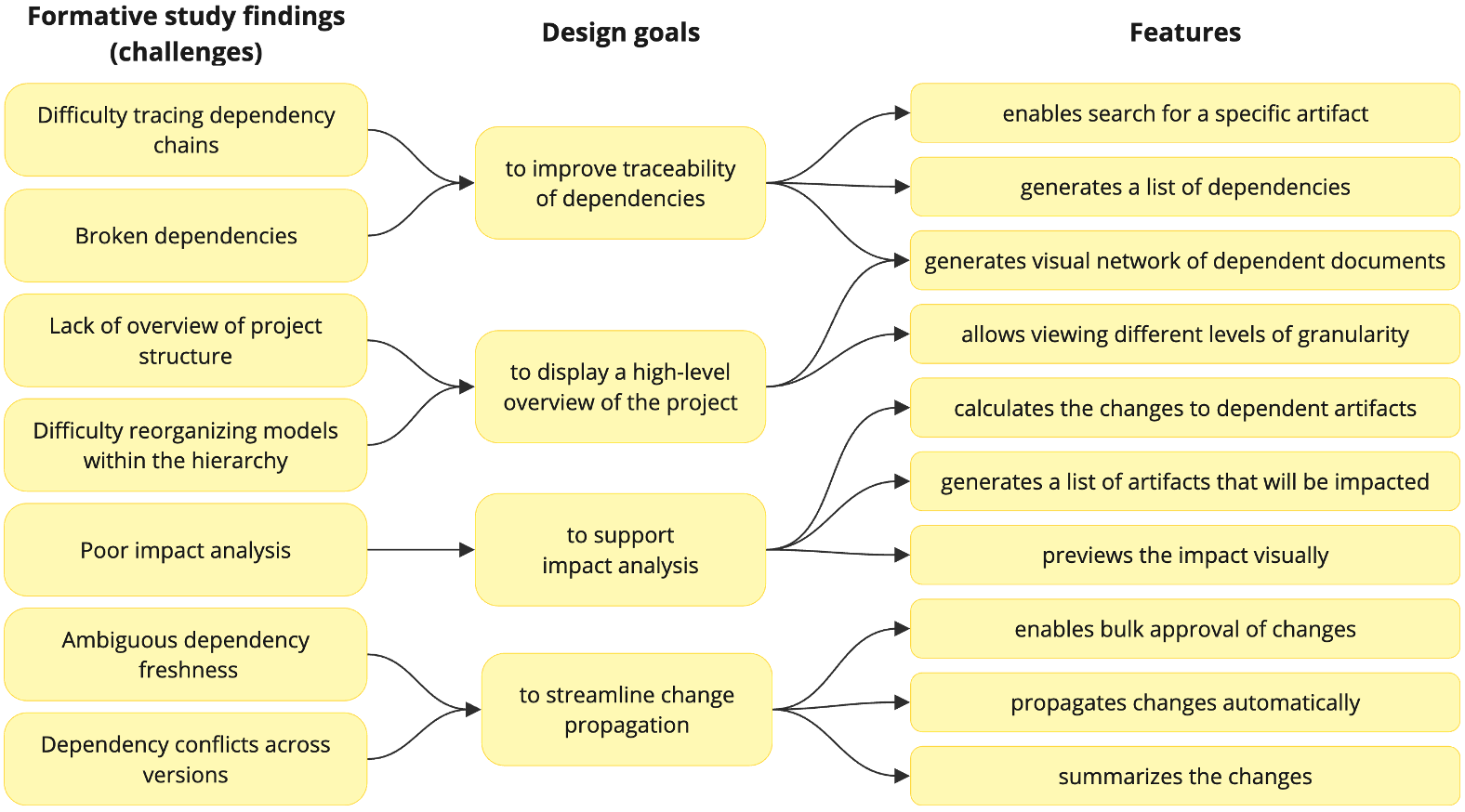}
  \caption{Overview of findings, design goals, and features for a dependency management tool.}
  \Description{Overview of findings, design goals, and features for a dependency management tool.}
  \label{designgoals}
\end{figure}

Our empirical investigations provide a comprehensive identification of dependency management challenges in cooperative CAD work. As the next step, we have distilled these findings into design goals and developed initial tool concepts to address the key challenges outlined in Section~\ref{results}. We focus on seven of the nine challenges identified, excluding the challenges of \textit{Disorganized design history} (\ref{chal:history}) and \textit{Messy navigation of the master sketch} (\ref{chal:mastersketch}). These two challenges each pertain to a very specific window in CAD  -- namely, the version control panel for design history, and the sketch interface for the master sketch -- and would benefit from specialized tool support integrated within these specific windows. We, therefore, leave these two challenges for future work, while focusing in this paper on the dependency management challenges that apply more broadly to external references. 
Based on our formative study findings, we outline four design goals and ten features for a CAD dependency management tool, summarized in Figure~\ref{designgoals} and discussed below.





\subsubsection{Goal 1: To improve traceability of dependencies}
CAD designers struggle to trace dependency chains (\ref{chal:tracing}), making it crucial for the tool to enhance awareness of how artifacts are linked. The tool should first allow users to search for a specific artifact (e.g., a document) and quickly locate it within the project. Next, the tool should generate a list of dependencies, showing both upstream (parent) and downstream (child) relationships. This list should be adjustable to show different levels of granularity, such as sketches, parts, and documents. By incorporating these features, designers can avoid the time-consuming task of manually searching across multiple documents. This improved traceability would also mitigate the risk of broken dependencies (\ref{chal:broken}), as users can confirm that no artifacts depend on a design before archiving or deleting it. 

\subsubsection{Goal 2: To display a high-level overview of the project}
Designers lack a clear overview of the project structure (\ref{chal:overview}), which makes it difficult to understand how documents relate to and impact each other. To support this need, the tool should generate a high-level visual network graph that maps the dependencies between documents. Users should be able to filter this view by level of granularity and expand or collapse nodes to enhance visibility. To further enhance awareness, users should also be able to select a specific document and visually trace the associated dependency chains. This visualization can also support decisions about document placement within the project hierarchy (\ref{chal:hierarchy}); for example, if a document appears high in the graph but has few or no dependents, it may prompt users to reconsider its placement.

\subsubsection{Goal 3: To support impact analysis}
Another challenge to target is the difficulty in identifying and understanding the effects of a change on related components (\ref{chal:change propagation}). Manually verifying that changes have been correctly propagated is a tedious and error-prone task. To address this, the tool should preview the changes to the user in two ways: (1) generating a text-based list of artifacts that will change and (2) visually previewing the impact in the network graph. Once the user makes the desired changes to the model, the tool should calculate and display the impact of these changes on the graph; for example, if the wheel diameter of a car is increased by 5 cm, the tool should update the position of the axle accordingly and indicate that the axle document has been modified. Impacted nodes should be highlighted, with nodes with errors or conflicts flagged (perhaps in red). This visual preview enables users to anticipate the changes before committing them. 

\subsubsection{Goal 4: To streamline change propagation}
The fourth goal addresses the challenges of ambiguous dependency freshness (\ref{chal:ambiguity}) and version conflicts (\ref{chal:conflict}). To mitigate these issues, the tool should support bulk approval of changes after reviewing the preview. Once approved, the tool should automatically propagate the changes, create new versions of the affected documents, and produce a summary report outlining all changes. This workflow ensures that design changes remain synchronized throughout the project. This idea is inspired by Team~Y's internal tool (described in Section~\ref{chal:ambiguity}), which detects a document's out-of-date references and enables batch version updates. Considering how critical this issue was for their team -- enough to warrant developing a custom solution -- we include this feature in our design goals. Our design concept builds on the foundation of Team~Y's tool by extending its functionality project-wide, aiming to provide similar value to other hardware development teams. Additionally, we propose generating change reports to provide transparency in the update process and automatic documentation, a long-standing pain point in CAD version control~\cite{marchenko_new_2011}.

\subsubsection{Tool Concept \& Scenario}
\begin{figure}[h]
  \centering
  \includegraphics[width=.87\linewidth]{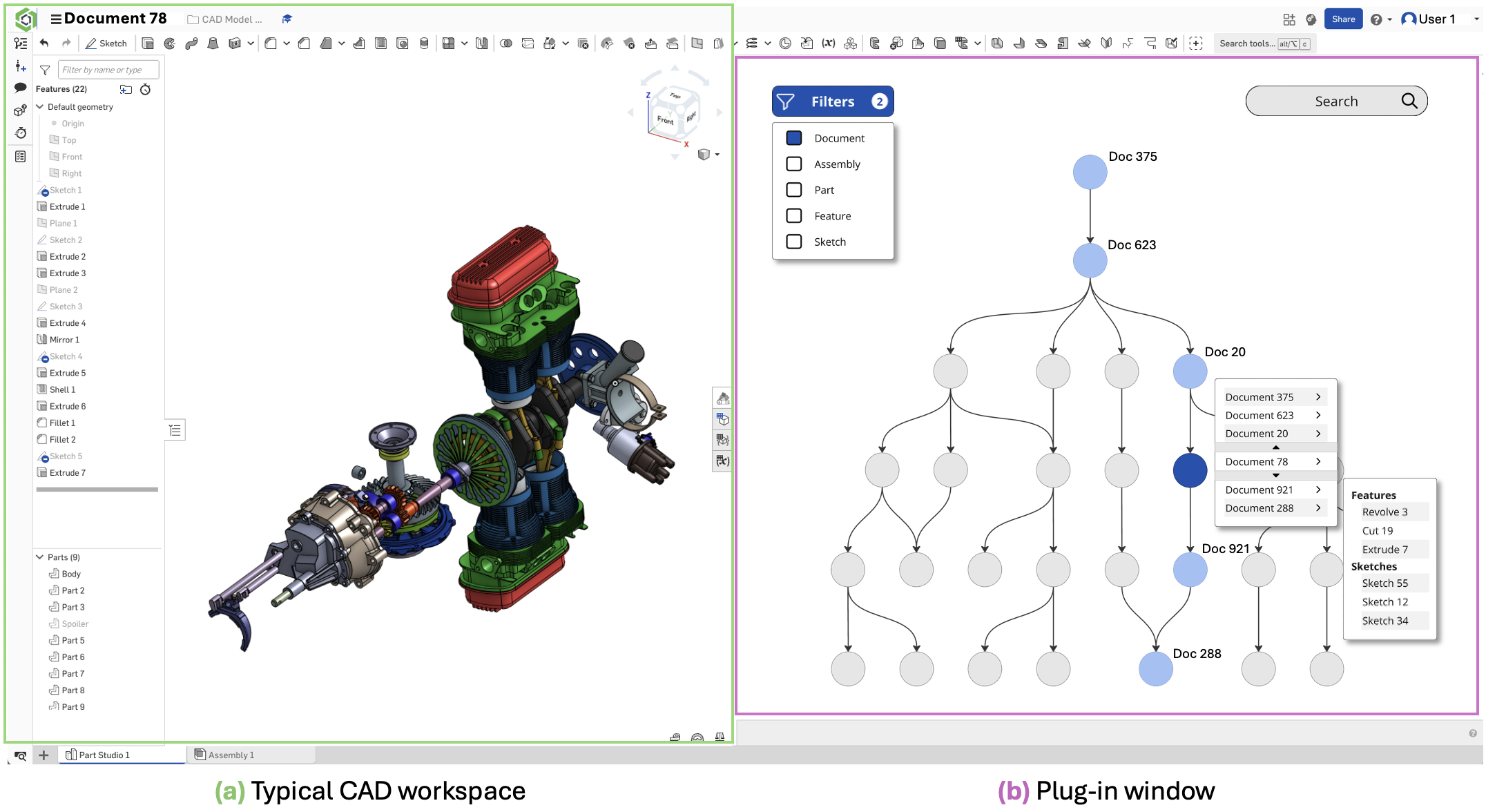}
  \caption{A tool can be implemented as a plug-in directly in the CAD platform interface. Note that these document names are dummy placeholders (e.g., ``Document 78''). CAD model adapted from~\cite{stark_classic_2024}.}
  \Description{Overview of findings, design goals, and features for a dependency management tool.}
  \label{mockup}
\end{figure}

The design goals and features that we proposed offer general implications for the builders of CAD platforms. However, to illustrate how such a tool could function, we developed a series of initial mockups. We propose a plug-in tool integrated into the CAD interface, allowing designers to easily access relevant dependency information within the context of their current work (see Figure~\ref{mockup}). In the plug-in window, a network graph maps the dependency relationships between documents, represented as nodes, with arrows pointing from parent to child. While all linked documents are displayed, the graph highlights the current open document (e.g., \textit{Document 78}) in dark blue, and all its parent and child documents in light blue for clear visibility. A pop-up window can also display a list of parent and child documents, with the ability to view more levels of granularity, such as the specific dependent features or sketches. 

The following user scenario illustrates how such a tool can improve CAD dependency management. Imagine Emily, a junior mechanical design engineer, preparing for a design sprint to improve the airflow of a cooling system. Emily has been assigned to redesign several components of the cooling system and integrate the changes without disrupting the overall product architecture.

\begin{figure}[b]
  \centering

  \begin{minipage}[t]{0.4\textwidth}
    \begin{subfigure}[b]{\textwidth}
      \centering
      \includegraphics[width=.9\textwidth]{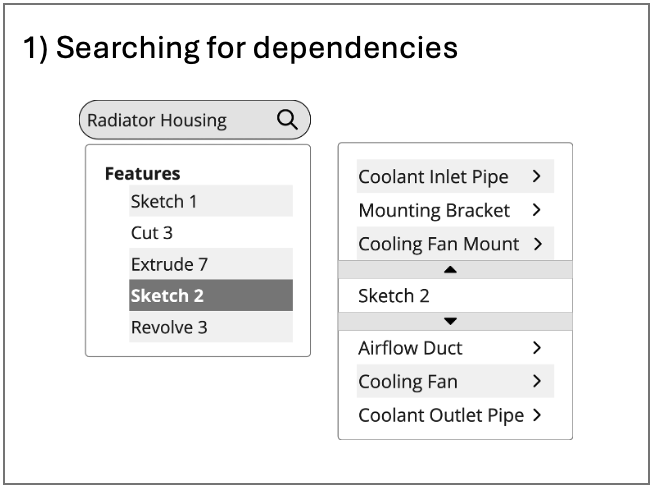}
      \caption{Search for a document's dependencies.}
      \label{scenarioA}
    \end{subfigure}

    \vspace{0.001\textheight}

    \begin{subfigure}[b]{\textwidth}
      \centering
      \includegraphics[width=.9\textwidth]{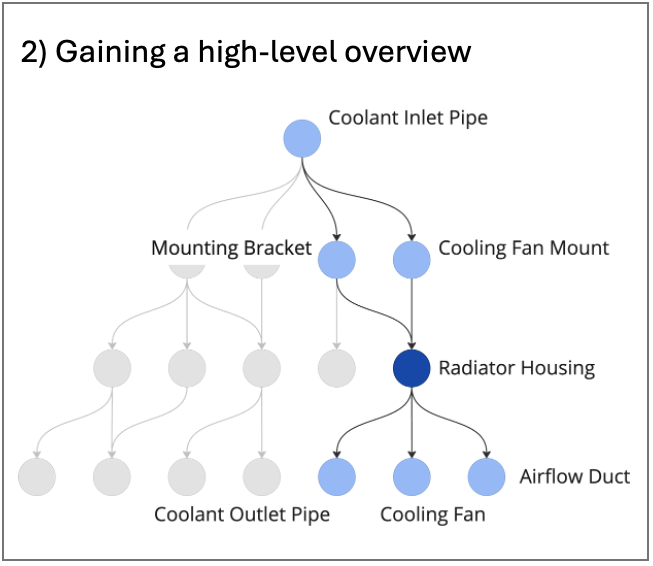}
      \caption{Visualize dependencies in a network.}
      \label{scenarioB}
    \end{subfigure}
  \end{minipage}
  \begin{minipage}[t]{0.58\textwidth}
    \begin{subfigure}[b]{\textwidth}
      \centering
      \includegraphics[width=.9\textwidth]{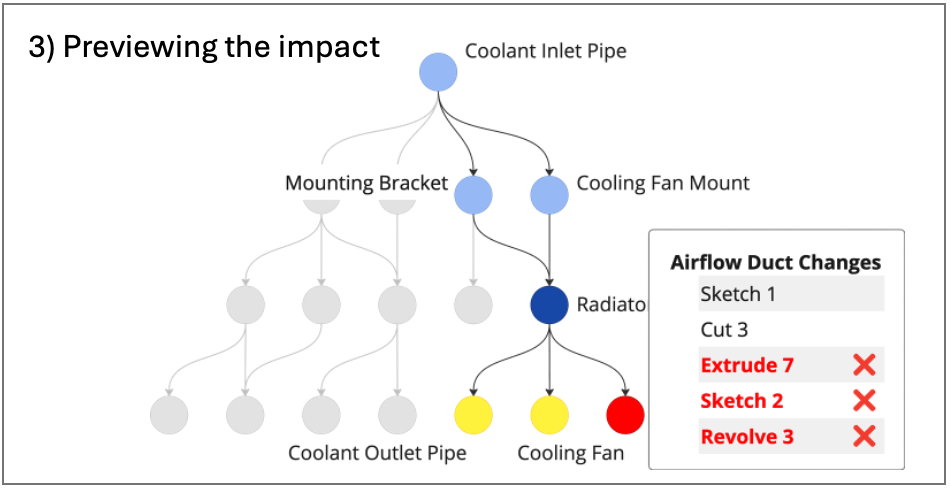}
      \caption{Preview the impact on dependent documents.}
      \label{scenarioC}
    \end{subfigure}

    \vspace{0.001\textheight}

    \begin{subfigure}[b]{\textwidth}
      \centering
      \includegraphics[width=.9\textwidth]{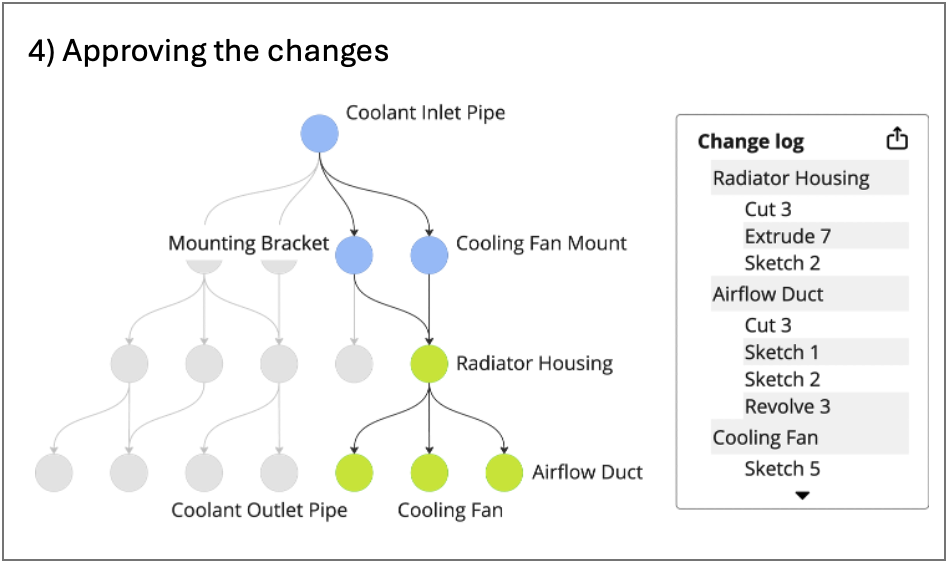}
      \caption{Bulk-approve the change propagation.}
      \label{scenarioD}
    \end{subfigure}
  \end{minipage}

  \caption{Example user scenario of a CAD dependency management tool.}
  \Description{Four subfigures illustrating different features: search, visualize, trace dependencies, and support change propagation.}
  \label{scenario}
\end{figure}

\paragraph{\textbf{Searching for dependencies}} Emily's first task is to update the design of the radiator housing. Aware that it depends on other parts of the cooling system, Emily avoids manually searching through numerous interconnected CAD files and folders, and instead uses the newly implemented dependency management tool. She types ``Radiator Housing'' into the tool's search bar (Figure~\ref{scenarioA}) and receives a list of upstream and downstream dependencies. The list not only provides direct links to relevant documents but also allows Emily to explore relationships at different levels of granularity, such as sketches and parts. This feature saves her from manually cross-referencing documents, giving her immediate clarity on which artifacts are connected.

\paragraph{\textbf{Gaining a high-level overview}} Emily needs to understand how changes to the Radiator Housing will affect other documents in the overall product. She uses the high-level overview feature of the tool (Figure~\ref{scenarioB}), which generates a visual network graph showing all documents and their dependency relationships within the cooling system project. Documents are represented by nodes, with arrows illustrating the dependencies. Using filters, Emily narrows her view to only the documents related to the Radiator Housing, revealing documents like the Airflow Duct and the Cooling Fan. Emily sees how modifications to the Radiator Housing may ripple through the cooling system. She expands nodes to explore specific sketches within the housing that might impact other documents, ensuring that no dependencies are overlooked.

\paragraph{\textbf{Previewing the impact}} Emily now modifies Radiator Housing. Before finalizing the changes, she activates the impact analysis feature (Figure~\ref{scenarioC}), to preview the changes. This generates a text-based list and a visual network graph that shows all affected artifacts, with nodes impacted by her changes highlighted in yellow. The Airflow Duct is flagged in red, signalling a potential conflict. The tool provides details on the flagged nodes, alerting her that her modifications interfere with several features (e.g., \textit{Extrude 7}, \textit{Sketch 2}). With this information, Emily can tweak her changes or consult a senior engineer to avoid introducing new problems. 

\paragraph{\textbf{Approving the changes}} After reviewing and resolving the design changes, Emily selects the option to bulk-approve the changes (Figure~\ref{scenarioD}). The tool then propagates the changes, creates new versions of all updated documents (highlighted in green), and generates a summary report, giving Emily a clear record of the changes made to the cooling system's design.


With this tool, Emily no longer needs to painstakingly trace dependencies, manually search through documents to find affected parts, or worry about unknowingly changing another designer's model. Although the design goals and features that we proposed here are by no means exhaustive, they target key challenges of CAD dependency management. Therefore, it is in the interest of CAD platform builders to implement such tools to enhance designers' awareness of dependencies and improve the efficiency of collaborative design.

\subsection{Modularity: A Proactive Dependency Management Approach}

Our study was motivated by the increasing complexity of hardware products, which comprise numerous components and even more interdependent relationships that are challenging to manage. We also proposed design goals for platform builders to help CAD designers better handle these dependencies. However, this is just one perspective to address the problem. We conjecture that managing dependencies can be approached: \textit{reactively}, by alleviating the challenges we identified, and \textit{proactively}, by adopting best practices that prevent challenges from arising in the first place.

Proactive management involves integrating strategies during the design process to avoid potential dependency issues. CAD researchers recommend minimizing the number of entities that depend on a single component to avoid unwanted changes~\cite{rosso_does_2022}, advising against overloading a single artifact with multiple dependencies. This best practice contradicts the master sketch philosophy, and may require a reevaluation of that architecture. For instance, designers could consider product modularity (the intentional decoupling of components~\cite{baldwin_design_1999}), by decomposing one highly intricate master sketch into several more manageable modules. In our interviews, we observed an emerging strategy that organizes master sketches in a hierarchy. At the top level, a global master sketch defines shared parameters, which are referenced by a set of mid-level master sketches, each of which governs a localized set of artifacts, creating a multi-layered hierarchy. This strategy extends prior CAD literature, which typically assumes a single skeleton structure~\cite{ciaccioli_investigation_2021}. While this modular approach can support more proactive dependency management, it also introduces a new challenge: designers must make deliberate decisions about where to draw boundaries between sketches. 

To improve modular design in CAD, future systems can draw from well-established software engineering principles. For example, object-oriented design patterns such as \textit{Facade}, \textit{Adapter}, and \textit{Mediator} from Gamma et al.'s catalog~\cite{gamma_design_1994} offer mechanisms to abstract internal complexity and define standardized points of interaction between components. Similar mechanisms in CAD could help isolate subsystems and make interdependencies more explicit and manageable. Currently, CAD models often depend on implicit geometric references (e.g., a line in one sketch defining the diameter of a part in another document), which are difficult to inspect or refactor. Analogous to software's \emph{interface segregation principle}, which encourages designing smaller, more focused interfaces between software components~\cite{martin_agile_2006}, CAD platforms could enforce clearer contracts at reference points -- e.g., requiring users to define named interface geometries rather than linking arbitrary sketch elements. While a promising direction, it must be adapted for hardware development, where parts can impact each other even without explicitly defined dependencies; this challenge is especially critical in assemblies with moving parts, where the motion path of one part can interfere with another in a separate module in ways that are difficult to predict or detect.

Additionally, CAD's top-down architecture resembles software's layered architectural styles, yet lacks the tools to manage such separation. Centralized ``master sketches'' are conceptually similar to central configuration modules in software, but CAD lacks automated tools for detecting violations of architectural boundaries or for restructuring modules when interdependencies grow tangled. Tools inspired by software's dependency injection and module boundaries~\cite{laigner_cataloging_2022,taelman_components_2022} could assist in refactoring CAD models as the design evolves. Our participants' frustrations with \textit{``chasing their own tail''} when resolving circular references underscore a need for dependency analyzers akin to those used in large software systems.

Importantly, modularity is not just a technical concern but a socio-technical one. In distributed teams, unclear dependencies often become coordination bottlenecks. As in software engineering, where Conway's Law implies that modular code reflects team communication structures~\cite{conway_committees_1968,herbsleb_splitting_1999}, CAD platforms could make modular boundaries more visible and enforceable to aid cross-team collaboration. By surfacing shared ``interfaces'' and ownership of different modules, CAD systems could better align technical architecture with organizational roles, reducing accidental overlap and misalignment across subsystems.

In summary, while modularity has long been a design ideal in software development, realizing it in CAD demands a rethinking of tooling support, grounded in technical patterns and software engineering collaborative practices. CSCW researchers are well-positioned to explore how these concepts translate to the CAD domain, examining how modularity affects collaboration and how teams negotiate, evolve, and contest modular boundaries in practice.

\subsection{Implications for Collaborative Work}\label{sec:implications}


Our work focused on the management of technical dependencies, which refers to the relationships between CAD artifacts. Designers must be aware of such dependencies in order to successfully create and modify designs. However, understanding technical dependencies is only the first step, and a natural progression is to explore how to manage the dependencies among people and teams~\cite{sosa_misalignment_2004}, i.e., \textit{work dependencies}. Below, we revisit the concepts of coordination and awareness to discuss the role of dependency management in collaborative design.

\paragraph{\textbf{Levels of collaborative activity.}}
Through the lens of Bardram's levels of collaborative activity~\cite{bardram_collaboration_1998}, we discuss how dependency management shapes coordination, cooperation, and co-construction.
Better awareness of technical dependencies supports \textbf{coordination} by clarifying which tasks depend on others, and how they should be accomplished (e.g., in series, in parallel)~\cite{eppinger_model_1994}. For example, project managers must determine which documents should be grouped together in a design sprint. Suppose subsystems A, B, and C are interdependent, and A and B are modified during the sprint; it makes sense also to include subsystem C. Redesigning related components concurrently reduces the likelihood of dependency conflicts and backward compatibility issues~\cite{venturini_depended_2023}. 

Beyond coordination, \textbf{cooperation} occurs when individuals focus on a shared design goal and make situated adjustments to their own and others' actions accordingly~\cite{bardram_collaboration_1998}. When designers know the technical dependencies associated with the CAD artifacts they are working on, it becomes easier to identify with whom they need to communicate~\cite{cataldo_socio-technical_2008}. For instance, if a designer knows that a colleague's part references a parameter they plan to change, they may proactively discuss the change, ensuring mutual understanding and reducing the risk of conflict~\cite{sosa_misalignment_2004}.

Finally, dependency management plays a role in \textbf{co-construction}, where collaborators jointly shape both the means and object of work~\cite{bardram_collaboration_1998}. We saw signs of co-constructive activity in our interviews, where participants described an emerging practice of multi-layered master sketches, and a custom tool that maintains dependency freshness within this hierarchy. Co-construction is both driven by, and a response to, increasingly complex technical dependencies. As CAD projects become more modularized, co-construction will involve reconceptualizing how tasks are distributed across teams -- shaping not only what is being designed but how the work itself is structured.

\paragraph{\textbf{Moving towards we-awareness.}}
As introduced in Section~\ref{awareness}, we-awareness is the socially recursive knowledge that collaborators have of each other~\cite{tenenberg_i-awareness_2016}, which is essential for successful distributed collaboration. Our proposed dependency management tool focuses on surfacing technical dependencies, which can support collaboration, as discussed above, but it does not fully address the need for we-awareness. A promising direction for future work is to incorporate social signals into the dependency graph -- for instance, indicating who is actively making changes to which documents, and the status of their changes (e.g., in progress, approved). By providing this kind of contextual information, the tool can enhance spatial awareness~\cite{schelble_i_2022} and facilitate the reciprocity needed for we-awareness.

\paragraph{\textbf{Dependency management in other collaborative domains.}}
In large-scale \textbf{software development} projects, developers often navigate intricate webs of dependencies. Changes in one module can cause unforeseen ripple effects across the system, harming maintainability and security. Our approach to visualizing dependencies and implementing proactive consistency checks can enhance tools like DepsRAG~\cite{alhanahnah_depsrag_2024}, a multi-agent framework designed to enhance developers' understanding of software dependencies using large language models. By integrating these strategies, software teams can achieve better traceability and mitigate risks associated with dependency changes.

Within \textbf{open-source ecosystems}, dependency management poses unique challenges. Many communities struggle with managing dependency risks, often overwhelmed by vulnerability alerts. Researching these alerts is resource-intensive, and attempting to address all of them can be even more costly~\cite{prana_out_2021,kharitonov_literature_2024,bifolco_on_2024}. Our findings suggest that integrating proactive consistency checks and dependency visualization tools can aid in prioritizing and addressing the most critical vulnerabilities, thereby enhancing the security and stability of open-source projects.

\textbf{Data scientists} face similar dependency awareness challenges during exploratory data analysis (EDA)~\cite{li_edassistant_2023,weinman_fork_2021}, particularly when trying to track dependencies between code cells and group them into coherent task-specific segments. Our findings on modularizing large, layered master sketches in CAD may offer transferable lessons -- such as creating hierarchical structures that could help contain the ``messiness''~\cite{head_managing_2019} of computational notebooks. Additionally, our study's emphasis on visualizing dependencies and implementing proactive consistency checks can significantly enhance collaborative EDA processes. For instance, tools like MLCask~\cite{luo_mlcask_2021} have been developed to manage component evolution in collaborative data analytics pipelines. By incorporating similar visualization and consistency management techniques, teams can improve the robustness and reproducibility of their workflows.


In conclusion, while our study primarily addressed technical dependencies in CAD, the implications of this work can help design teams better plan activities and facilitate collaborative efforts. Although our focus is on CAD, these implications also shed light on awareness and coordination needs in CSCW domains as a whole, where complex dependencies are common in fields like software development~\cite{treude_awareness_2010} or data science~\cite{zhang_how_2020,mao_how_2019}.

\subsection{Limitations \& Future Work}

This paper focuses on understanding the user requirements for managing technical dependencies between CAD artifacts, but as discussed in Section~\ref{sec:implications}, this does not encompass all the dependency needs in cooperative CAD work, such as work dependencies. Nonetheless, we believe that our systematic investigation of dependency management challenges is a necessary first step, and future work will extend this investigation to explore socio-technical dependencies. 

Next, like other empirical studies that conducted formative research with participants from a single organization~\cite{oleary_charrette_2018,khadpe_discern_2024}, our findings have limited generalizability. However, to mitigate this limitation, we included insights from online forum discussions to broaden the scope of perspectives represented in our dataset. 

Our study is also limited by its focus on a single CAD platform, Onshape. We intentionally chose Onshape because of its advanced support for external references, but we cannot claim that the challenges we identified are generalizable to all CAD software. Nevertheless, selecting Onshape was essential, allowing us to investigate persistent challenges even within a state-of-the-art system.

Finally, in this work, we focused on a two-phase formative study of engineers' experiences with CAD dependency management, finding several fruitful user needs. Based on these findings, we developed four design goals for a dependency management tool to address these challenges and proposed initial tool concepts. It must be recognized that these design goals are not exhaustive, and the features we present do not represent a comprehensive checklist for CAD dependency management. For future work, we will develop and implement a functional tool based on the concepts proposed in this paper. We will evaluate this tool using one or more methods, such as usability studies~\cite{weinman_fork_2021,wittenhagen_chronicler_2016}, design walkthroughs~\cite{han_passages_2022}, or longitudinal studies~\cite{sedlmair_cardiogram_2011}, to further refine the design and assess its effectiveness.

\section{Conclusion}
In this work, we conducted two empirical formative studies, seeking to better understand CAD dependency management challenges. Through a thematic analysis of 100 popular user discussions in online CAD forums and semi-structured interviews with 10 professional designers, we uncovered nine key challenges that hinder workspace awareness and effective coordination within hardware development teams. In an effort to enhance CAD dependency management, we distilled these challenges into design goals and corresponding features that are essential for a CAD dependency management tool. Beyond challenges and design goals, we also contribute initial tool concepts that could implement these features as a plug-in for CAD platform interfaces. Our findings and proposed solutions lay the groundwork for future development and evaluation of tools that can better support designers in managing CAD dependencies and improving collaboration in the hardware development domain.

\begin{acks}
We acknowledge the support of the Government of Canada’s New Frontiers in Research Fund (NFRF), [NFRFE-2022-00543]. We also thank our interview participants for generously sharing their time and valuable insights.
\end{acks}

\bibliographystyle{ACM-Reference-Format}
\bibliography{sample-base}

\received{October 2024}
\received[revised]{April 2025}
\received[accepted]{August 2025}

\end{document}